\documentclass[conference]{IEEEtran}

\usepackage{graphicx}
\usepackage{algorithmic,algorithm}
\usepackage{amsfonts}
\usepackage{multirow}
\usepackage[cmex10]{amsmath}
\begin{document}
%
\title{A Low-Delay Low-Complexity EKF Design for Joint Channel and CFO Estimation in Multi-User Cognitive Communications}

\author{\authorblockN{Pengkai Zhao}
        \authorblockA{Electrical Engineering, UCLA, CA, USA}
        \and
        \authorblockN{Cong Shen}
        \authorblockA{Qualcomm Inc., San Diego, CA, USA}
        }

%
%
%


\maketitle

\begin{abstract}
Parameter estimation in cognitive communications can be formulated as a multi-user estimation problem, which is solvable under maximum likelihood solution but involves high computational complexity. This paper presents a time-sharing and interference mitigation based EKF (Extended Kalman Filter) design for joint CFO (carrier frequency offset) and channel estimation at multiple cognitive users. The key objective is to realize low implementation complexity by decomposing high-dimensional parameters into multiple separate low-dimensional estimation problems, which can be solved in a time-shared manner via pipelining operation. We first present a basic EKF design that estimates the parameters from one TX user to one RX antenna. Then such basic design is time-shared and reused to estimate parameters from multiple TX users to multiple RX antennas. Meanwhile, we use interference mitigation module to cancel the co-channel interference at each RX sample. In addition, we further propose adaptive noise variance tracking module to improve the estimation performance. The proposed design enjoys low delay and low buffer size (because of its online real-time processing), as well as low implementation complexity (because of time-sharing and pipeling design). Its estimation performance is verified to be close to Cramer-Rao bound.
\end{abstract}


%
\IEEEpeerreviewmaketitle

\section{Introduction}
Cognitive communication system is widely accepted as a perspective way in increasing the spectrum efficiency of wireless networks, where primary links and secondary links can usually co-exist in the network, resulting in an interference limited environment. Parameter estimation in cognitive communications is a challenging problem because of (i) the existence of co-channel interference, and (ii) the high-dimensional parameters from multiple TX users to multiple RX antennas. In particular, note that different TX users often have independent carrier frequency offset (CFO) values (including both oscillator offsets and doppler offsets), which usually introduce serious nonlinear components within the observed signal, complicating the estimation problem. Meanwhile, channel responses from multiple TX users to multiple RX antennas can result in a set of high-dimensional parameters, which are also difficult to estimate. Finally, due to the existence of multi-user interference, CFO and channel parameters usually have to be treated together and be estimated in a joint way so as to approach the optimal performance, which further increases the estimation complexity.

Without loss of generality, this paper assumes Orthogonal Frequency Division Multiplexing (OFDM) system, which is an overwhelming choice for modern wireless systems. The classical CFO and channel estimation method in a single-user OFDM system is based on two repeated training symbols \cite{OFDM}. It has low implementation complexity and near-optimal performance, but only applies to a single-user scenario\footnotemark. In multi-user OFDMA systems with unique subcarrier set per TX user, CFO and channel parameters can be recovered by exploiting distinct subcarrier structures among TX users (\cite{OFDMA_1, OFDMA_2}). But this method requires separate subcarrier allocation for different users. Consequently, in a general multi-user cognitive system without specific subcarrier allocation per user and with overlapped training symbols, ML and EM related methods seem to be the only applicable choice, where all TX users' parameters have to be formulated into a maximum likelihood (ML) estimation problem \cite{ML}, which is solvable under Expectation Maximization (EM) method \cite{EM} in an iterative way. However, since the entire OFDM block is stored offline and is iteratively processed multiple times, these ML and EM approaches often require high computational complexity and high processing delay.

\footnotetext{It is also applicable for multi-user scenario with non-overlapping training symbols, but this is not the case considered in this paper.}

Based on above considerations, this paper will focus on using Kalman filter structure to estimate the CFO and channel parameters in multi-user cognitive communications. Our major objective is to achieve low-complexity and low-delay estimation performance in cognitive systems. In general, Kalman filter is a good candidate for low delay and low complexity parameter estimation primarily due to its real-timing processing property. It has been conventionally used for CFO and channel estimation in multi-user OFDM systems, e.g., the FFT-Block EKF design in \cite{Kalman_MIMO}, the parallel EKF design in \cite{Kalman_Particle}, and the particle filter design in \cite{Kalman_Particle}. However, these existing designs inherently suffer from multiple issues related to complexity, delay and buffer size as follows:
\begin{enumerate}
\item Block EKF design in \cite{Kalman_MIMO} operates on an FFT-block basis, which grows increasingly complex as FFT size becomes large (e.g., 2048 FFT size). Also, parameters estimated in this method are handled in a high dimension manner.
\item Parallel EKF design in \cite{Kalman_Particle} also operates on an FFT-block basis, marking it complex under large FFT size. Parameters in this design are jointly estimated by calculating the covariance information between different users, leading to a high matrix dimension.
\item Beyond FFT size and parameter dimension issues similar to item 1 and 2, particle filter design in \cite{Kalman_Particle} needs to repeat the Kalman operation at multiple particle samples, yielding a multiplicative effect on complexity.
\end{enumerate}
To summarize, the major challenge in implementing a low-complexity EKF design lies in the factors of: (i) multiple TX users; (ii) multiple RX antennas; (iii) high parameter dimension; and (iv) large FFT size (e.g., 2048 size).

With low-complexity and low-delay requirement in mind, this paper will present a time-sharing and interference mitigation based Extended Kalman Filter (EKF) design for multi-user cognitive communications, which can estimate the CFO and channel parameters from multiple TX users to multiple RX antennas in a time-sharing manner. Here low delay property is achieved by using Kalman filter estimation at each RX sample in a real-time manner, and low complexity property is achieved by reusing a single user EKF design in a time-sharing{\footnotemark} and pipeling way. We first present a fundamental EKF design that estimates the CFO and channel parameters from one TX user to one RX antenna. Then such basic EKF design is reused in a time-shared way to estimate the parameters from multiple TX users to multiple RX antennas. Meanwhile, at each RX sample, an interference mitigation strategy is developed to estimate and remove the expected multi-user interference. In addition, we provide an adaptive noise variance tracking module to further enhance the estimation performance. Because of the usage of EKF structure, our design is essentially different from the particle filters in \cite{Kalman_Particle} and the EM method in \cite{EM}. Our design is also different from the Parallel-EKF design in \cite{Kalman_Particle} and the FFT-Block EKF design in \cite{Kalman_MIMO} at the following perspectives: (i) our design runs at each time domain RX sample, not at an FFT-block basis; (ii) our design treats each user separately, not jointly; (iii) our design can be implemented in a time-sharing way, which is less considered in \cite{Kalman_MIMO} and \cite{Kalman_Particle}; (iv) system model in our design is different from the ones in \cite{Kalman_MIMO, Kalman_Particle} by integrating CFO parameter into channel response (see Eqn. (\ref{state_chan}) in section II). Analysis and simulations results validate that our proposed design can closely approach the Cramer-Rao bound, and has lower computational complexity than the ones in \cite{Kalman_MIMO, Kalman_Particle}. Finally, although cognitive communication is a typical application scenario for our proposed design, it is also applicable in many other multi-user systems that satisfy the conditions presented in section II.

\footnotetext{Time-sharing in this paper indicates that the same hardware module can be reused by different processes at separate time slots.}

\section{System Model}

\subsection{Problem Formulation}
We consider a total of $Q$ TX users in the cognitive network. One of them is the primary TX user (i.e., base station), and the rest are all secondary TX users. Primary TX user's transmission is based on a time division MAC protocol, where time is divided into different time frames with equal duration. Secondary users can maintain time synchronization with the primary TX user by learning and synchronizing with its time frames. Each secondary RX user is equipped with $N_A$ multiple antennas to decode the packets. Without loss of generality, we assume that every TX user has only one spatial stream, and there exists $Q \leq N_A$. Also, every TX user has a distinct training symbol{\footnotemark} $s_q(n)$ with $1 \leq q \leq Q$ and $0 \leq n \leq N_{F}-1$. Here $N_{F}$ is the FFT size of OFDM system.

\footnotetext{This unique training symbol can be determined according to either the unique user ID in the network, or the access order in the current time frame.}

Each TX user has an independent carrier frequency offset (CFO) that is caused by both the oscillator offset and the doppler offset. Denote TX user $q$'s CFO value as $\varepsilon_{q}$. For a given secondary RX user, the channel from TX user $q$ to the $m$th RX antenna of this secondary user is denoted as $h_{q,m}(p_{l}^{q,m})$, $\forall\ 1 \leq l \leq L_{\rm max}$. Here $L_{\rm max}$ is the number of time domain paths in the channel response, and $p_{l}^{q,m}$ is an integer value representing the relative delay of the $l$th path. We assume that all $p_{l}^{q,m}$ values have already been determined at an early stage (e.g., using PN sequences at coarse synchronization). The received signal at the $m$th RX antenna is derived as:
\begin{IEEEeqnarray}{rCl}
y_m(n)&=&\sum_{q=1}^Q \exp\left(j\frac{2\pi\varepsilon_{q} n}{N_{F}}\right) \sum_{l=1}^{L_{\rm max}}h_{q,m}(p_{l}^{q,m}) s_{q}\left[(n-p_{l}^{q,m})_{N_{F}}\right]\nonumber\\
&&+z_m(n), 0 \leq n \leq N_F-1
\end{IEEEeqnarray}
where $(n-p_{l}^{q,m})_{N_{F}}=\left\{ (n - {p_{l}^{q,m}}) {\ \rm mod\ } N_{F}\right\}$ is circular shift, and $z_m(n)$ is the background noise at the $m$th RX antenna.

The task in this work is to estimate CFO parameter $\varepsilon_q$ and channel parameter $h_{q,m}(p_{l}^{q,m})$ for all users ($1 \leq q \leq Q$) and all antennas ($1 \leq m \leq N_A$). Obviously, the optimal estimation is the solution to this maximum likelihood (ML) problem:
\begin{IEEEeqnarray}{rCl}
\label{ML}
&&\min \sum_{m=1}^{N_A} {\big{|}}y_m(n) - \nonumber\\
&&\sum_{q=1}^Q \exp\left(j\frac{2\pi{\widehat{\varepsilon}}_{q} n}{N_{F}}\right) \sum_{l=1}^{L_{\rm max}}{\widehat h}_{q,m}({p_{l}^{q,m}}) s_{q}\left[(n-p_{l}^{q,m})_{N_{F}}\right]{\big{|}}^2
\end{IEEEeqnarray}
where ${\widehat{\varepsilon}}_{q}$ and ${\widehat h}_{q,m}({p_{l}^{q,m}})$ represent the estimated values. There are a total of $(L_{\rm max}N_A+1)Q$ parameters in Eqn. (\ref{ML}), which constitutes a high-dimensional parameter estimation problem.

\subsection{State-Space Formulation}
ML solution can generally approach the optimal performance but it requires huge computations, which are highly undesirable in most systems. Instead, this paper proposes an EKF design for the estimation of the CFO and channel parameters, which can sequentially update the estimation results at each RX sample, resulting in low buffer size and low estimation delay. Initially, it is straightforward to directly apply an EKF design at Eqn. (\ref{ML}) by building all CFO and channel parameters into one state vector, whose dimension is as high as $(L_{\rm max}N_A+1)Q$. This method will significantly increase the complexity of the derived Kalman filter. With such complexity consideration in mind,  we first propose a low-dimensional EKF design that can estimate the parameters from one TX user to one RX antenna, which has only $(L_{\rm max}+1)$ parameters. Then we reuse this fundamental EKF design in a time-shared manner to estimate the parameters from multiple TX users to multiple RX antennas. In this way, high-dimensional parameters are estimated by sequentially reusing a low-dimensional estimator, which reduces the complexity of the proposed EKF design.

We first present an RX signal formulation from the perspective of TX user $q$ and the $m$th RX antenna as:
\begin{IEEEeqnarray}{rCl}
\label{rx_single}
y_{q,m}(n)&=&\exp\left(j\frac{2\pi\varepsilon_{q} n}{N_{F}}\right) \sum_{l=1}^{L_{\rm max}}h_{q,m}(p_{l}^{q,m}) s_{q}\left[(n-p_{l}^{q,m})_{N_{F}}\right]\nonumber\\
&&+z_{q,m}(n),
\end{IEEEeqnarray}
here $y_{q,m}(n)$ is extracted from $y_{m}(n)$ with the aid of interference mitigation module, and $z_{q,m}(n)$ represents the residual noise at TX user $q$ and the $m$th RX antenna, which includes both the residual co-channel interference from other TX users and the background noise at the $m$th RX antenna. Additionally, the initial value of $y_{q,m}(n)$ without any interference mitigation is set to $y_{q,m}(n) = y_{m}(n)$. Details about the interference mitigation module will be given in section III.

Now we define the associated state vector as:
\begin{IEEEeqnarray}{rCl}
&&{\rm {\bf X}}_{q,m}(n) = \left[ \varepsilon_q, {\rm {\bf H}}_{q,m}(n) \right]^T, \\
\label{state_chan}
&&{\rm {\bf H}}_{q,m}(n) = \exp\left(j\frac{2\pi n\varepsilon_q}{N_F}\right)\left[ h_{q,m}(p^{q,m}_{1}), ..., h_{q,m}(p^{q,m}_{L_{\rm max}})\right].
\end{IEEEeqnarray}
The state-space model for TX user $q$ and the $m$th RX antenna can be derived from (\ref{rx_single}) as:
\begin{IEEEeqnarray}{rCl}
&&{\rm {\bf X}}_{q,m}(n) = f\left\{ {\rm {\bf X}}_{q,m}(n-1) \right\}={\rm {\bf D}}_{q,m}(\varepsilon_q){\rm {\bf X}}_{q,m}(n-1), \\
&&{\rm {\bf D}}_{q,m}(\varepsilon_q) = \left[ \begin{array}{cc}
 1 & {\rm {\bf 0}}_{1\times L_{\rm max}} \\
{\rm {\bf 0}}_{L_{\rm max} \times 1} & \exp\left(j 2\pi \varepsilon_q/N_{ F}\right) {\rm {\bf I}}_{L_{\rm max}\times L_{\rm max}}
\end{array} \right]\\
&&y_{q,m}(n) = {\rm {\bf G}}_{q,m}(n){\rm {\bf X}}_{q,m}(n) + z_{q,m}(n), \\
&&{\rm {\bf G}}_{q,m}(n) = \big[ 0, s_q\left[(n-p^{q,m}_{1})_{N_{F}}\right], s_q\left[(n-p^{q,m}_{2})_{N_{F}}\right], ..., \nonumber\\ &&s_q\left[(n-p^{q,m}_{L_{\rm max}})_{N_{F}}\right] \big].
\end{IEEEeqnarray}
Finally, it is worth mentioning that the derived state-space formulation is a nonlinear model, since there is a nonlinear component $\exp\left(j 2\pi \varepsilon_q/N_F\right)$ in the matrix ${\rm {\bf D}}_{q,m}(\varepsilon_q)$.

\section{Interference Mitigation based EKF Design}
This section sequentially describes (i) the basic EKF design that aims at only one TX user and one RX antenna, (ii) the interference mitigation module that cancels co-channel interference at each RX sample, (iii) the proposed adaptive noise variance tracking module, and (iv) the overall paradigm of the proposed design.

\subsection{Fundamental EKF Design}
The key idea behind the EKF design is using Jacobian derivation to linearize the nonlinear matrix ${\rm {\bf D}}_{q,m}(\varepsilon_q)$ at local estimates:
\begin{IEEEeqnarray}{rCl}
\label{Jacobian}
&&{\rm {\bf F}}_{q,m}(n-1)= \left. \frac{\partial f\left( {\rm {\bf X}}_{q,m}(n-1) \right)} {\partial {\rm {\bf X}}_{q,m}(n-1) }  \right|_{\widehat{{\rm {\bf X}}}_{q,m}(n-1|n-1)} \nonumber \\
&&= \left[ \begin{array}{cc}
 1 & {\rm {\bf 0}}_{1\times L_{\rm max}} \\
\alpha(n-1){\widehat{\rm {\bf H}}}_{q,m}^T(n-1|n-1) & \exp\left(\alpha(n-1)\right) {\rm {\bf I}}_{L_{\rm max}\times L_{\rm max}}
\end{array} \right] \nonumber\\
\\
&&\alpha(n-1) = j 2\pi \widehat{\varepsilon}_{q,m}(n-1|n-1)/{N_{ F}}
\end{IEEEeqnarray}
where ${\widehat{{\rm {\bf X}}}_{q,m}(n-1|n-1)}$ represents the estimated state vector after processing the $(n-1)$th RX sample. Based on (\ref{Jacobian}), the prediction steps in our fundamental EKF design are:
\begin{IEEEeqnarray}{rCl}
&&{\widehat{{\rm {\bf X}}}_{q,m}(n|n-1)} =\nonumber\\
&&{\rm {\bf D}}_{q,m}( \widehat{\varepsilon}_{q,m}(n-1|n-1) ){\widehat{{\rm {\bf X}}}_{q,m}(n-1|n-1)}, \\
&&{\rm {\bf P}}_{q,m}(n|n-1) =\nonumber\\
&&{\rm {\bf F}}_{q,m}(n-1){\rm {\bf P}}_{q,m}(n-1|n-1){\rm {\bf F}}_{q,m}^H(n-1).
\end{IEEEeqnarray}
And the updating steps are as follows:
\begin{align}
\label{variance}
&{\rm {\bf K}}_{q,m}(n) = {\rm {\bf P}}_{q,m}(n|n-1){\rm {\bf G}}_{q,m}^H(n)\times \nonumber\\
&\left[  {\rm {\bf G}}_{q,m}(n){\rm {\bf P}}_{q,m}(n|n-1){\rm {\bf G}}_{q,m}^H(n) + \sigma_{q,m}^2(n) \right]^{-1},\\
&{\rm {\bf P}}_{q,m}(n|n) = \nonumber\\
&\left[ {\rm {\bf I}} - {\rm {\bf K}}_{q,m}(n){\rm {\bf G}}_{q,m}(n) \right] {\rm {\bf P}}_{q,m}(n|n-1), \\
\label{new_info}
&{\widehat{{\rm {\bf X}}}_{q,m}(n|n)} = {\widehat{{\rm {\bf X}}}_{q,m}(n|n-1)} + \nonumber\\
&{\rm {\bf K}}_{q,m}(n)\left[ y_{q,m}(n) - {\rm {\bf G}}_{q,m}(n){\widehat{{\rm {\bf X}}}_{q,m}(n|n-1)}\right].
\end{align}
Here $\sigma_{q,m}^2(n)$ represents the variance of the observation noise $z_{q,m}(n)$ in Eqn. (\ref{rx_single}). Also, CFO estimate $\widehat{\varepsilon}_{q,m}(n|n)$ in the state vector ${\widehat{{\rm {\bf X}}}_{q,m}(n|n)}$ should only use its
real part as $\widehat{\varepsilon}_{q,m}(n|n) = {\rm Real}\left[\widehat{\varepsilon}_{q,m}(n|n)\right]$. Finally, the EKF design presented above is only used for the parameter estimation of one TX user and one RX antenna. This basic EKF design is then iterated in a time-shared manner to estimate the parameters of all TX users ($1 \leq q \leq Q$) and all RX antennas ($1 \leq m \leq N_A$).

\subsection{Interference Mitigation and Refined CFO Estimation}

Before describing the interference mitigation module, we first look at the refinement of the CFO estimates. Although TX user $q$ has only one CFO parameter, our proposed EKF design can result in $N_A$ different estimates that are derived from $N_A$ RX antennas, which are denoted as $\widehat{\varepsilon}_{q,m}(n|n-1)$, $1 \leq m \leq N_A$. As a result, we can use these $N_A$ different estimates to get a refined result $\widehat{\varepsilon}_{q}(n|n-1)$, which is calculated as:
\begin{IEEEeqnarray}{rCl}
&&\widehat{\varepsilon}_{q}(n|n-1)=\nonumber\\
&&\sum_{m=1}^{N_A}\frac{1/{\rm {\bf P}}_{q,m}(n|n-1)_{(1,1)}}{\sum_{r=1}^{N_A}1/{\rm {\bf P}}_{q,r}(n|n-1)_{(1,1)}}\widehat{\varepsilon}_{q,m}(n|n-1),
\end{IEEEeqnarray}
where ${\rm {\bf P}}_{q,m}(n|n-1)_{(1,1)}$ denotes ${\rm {\bf P}}_{q,m}(n|n-1)$'s element located at the $1^{\rm st}$ row and $1^{\rm st}$ column.

Recall that $y_{q,m}(n)$ involved in (\ref{rx_single}) and (\ref{new_info}) is extracted from the original RX signal $y_{m}(n)$ with the help of an interference mitigation strategy. Having derived the refined CFO estimate $\widehat{\varepsilon}_{q}(n|n-1)$, now the interference mitigation process can be applied at $y_{q,m}(n)$ as follows:
\begin{IEEEeqnarray}{rCl}
&&y_{q,m}(n) = y_{m}(n) - \nonumber\\
&&\sum_{u=1,u\neq q}^{Q} \exp\left(j\frac{2\pi\widehat{\varepsilon}_{u}(n|n-1)}{N_{F}}\right)\cdot {\rm {\bf G}}_{u,m}(n){\widehat{{\rm {\bf X}}}_{u,m}(n|n-1)}.\nonumber\\
\end{IEEEeqnarray}


\subsection{Adaptive Noise Variance Tracking}
Since $y_{q,m}(n)$ is extracted from $y_{m}(n)$ via interference mitigation module, the variance of $z_{q,m}(n)$, (i.e., $\sigma_{q,m}^2(n)$ used in Eqn. (\ref{variance})), is  varying during the convergence process of the interference mitigation module, which has to be adaptively tracked. Such variance tracking is based on the following observation:
\begin{IEEEeqnarray}{rCl}
&&\mathbb{E}\left| y_{q,m}(n) - {\rm {\bf G}}_{q,m}(n){\widehat{\rm {\bf X}}}_{q,m}(n|n-1) \right|^2 \approx \nonumber\\
&&{\rm {\bf G}}_{q,m}(n){\rm {\bf P}}_{q,m}(n|n-1){\rm {\bf G}}^H_{q,m}(n) + \sigma_{q,m}^2(n).
\end{IEEEeqnarray}
Using (19), noise variance $\sigma_{q,m}^2(n)$ can be tracked as:
\begin{IEEEeqnarray}{rCl}
\sigma_{q,m}^2(n) &=& \left[ 1 - \frac{1-b}{1-b^{n+1}} \right]\cdot \sigma_{q,m}^2(n-1) + \nonumber\\
&&\frac{1-b}{1-b^{n+1}}\cdot \left\{ \max\left[e_{q,m}(n),0\right]\right\}, \\
e_{q,m}(n) &=& \left| y_{q,m}(n) -
{\rm {\bf G}}_{q,m}(n){\widehat{\rm {\bf X}}}_{q,m}(n|n-1) \right |^2 - \nonumber\\
&&{\rm {\bf G}}_{q,m}(n){\rm {\bf P}}_{q,m}(n|n-1){\rm {\bf G}}^H_{q,m}(n),
\end{IEEEeqnarray}
where $b=0.99$ is the decay factor used to exponentially weight the history values.

\subsection{Block Diagram}
\begin{figure}
\centering
\includegraphics[width=2.5in]{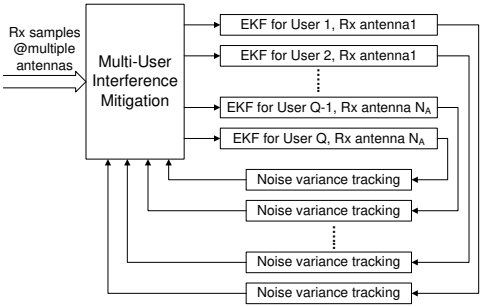}
\caption{Block diagram of our proposed design.} \label{Diagram}
\end{figure}

The complete functional diagram of our proposed design is shown in Fig. \ref{Diagram}. In this paradigm, received samples at all RX antennas are first processed in the interference mitigation module. Then the resultant samples are sequentially processed in the basic EKF module and noise variance tracking module. For the ease of description, all components in Fig. \ref{Diagram} are depicted in a parallel manner. However, in practice these design components can be implemented in a time-shared manner, and only one single EKF module is physically required.

\section{Simulations and Discussions}

\subsection{Parameters Setup}
The OFDM system built in the simulation has a bandwidth of 20MHz and a FFT size $N_F=2048$. Each TX user's CFO value is independently and randomly generated within the range of [-2, 2]\footnotemark. Wireless channels are generated using the SUI-3 channel model \cite{SUI_Model}, which has $L_{\rm max}=3$ non-zero paths at the time domain. SNR in this paper is defined as the ratio of the signal power to the noise power at one RX antenna, i.e., ${\rm SNR} = \sigma_R^2/\sigma_Z^2$ where $\sigma_R^2$ is the total received signal power at one RX antenna that is coming from all TX users, and $\sigma_Z^2$ is the power of the background noise. In the simulation, CFO and channel parameters are estimated using one OFDM training symbol with $N_F=2048$ samples. Estimation results are validated via the mean square error (MSE) performance. Specifically, MSE for channel estimation is defined as a normalized version:
\begin{eqnarray}
{\rm MSE}(h_{q,m})=\frac{\sum_{l=1}^{L_{\rm max}}\left|{\widehat h}_{q,m}(p_{l}^{q,m}) - h_{q,m}(p_{l}^{q,m})\right|^2}{\sum_{l=1}^{L_{\rm max}}\left|h_{q,m}(p_{l}^{q,m})\right|^2}.
\end{eqnarray}
Cramer-Rao bounds for the MSE results of CFO and channel estimation can be derived according to \cite{CR_bound} as:
\begin{eqnarray}
{\rm CRB}_{\rm CFO}({\rm SNR}) = \frac{3Q}{2\pi^2\cdot N_F \cdot {\rm SNR} \cdot N_A }, \\
{\rm CRB}_{\rm Chan}({\rm SNR}) = \left( \frac{L_{\rm max}}{N_F}+\frac{3}{2N_F} \right)\frac{Q}{{\rm SNR}}.
\end{eqnarray}

\footnotetext{Since integer frequency offsets can generally be estimated during the coarse synchronization stage, CFO value at fine synchronization stage is usually between -0.5 and 0.5. But here we use range 2 to demonstrate our design's estimation performance.}

\newcommand{\tabincell}[2]{\begin{tabular}{@{}#1@{}}#2\end{tabular}}
\begin{table}
\caption{Complexity Comparison (Number of Complex Multiplications)} \centering \label{Table_complexity}
\begin{tabular}{|c|c|}
\hline \textbf{Design Name} & \textbf{Number of Complex Multiplications}\\
\hline \textrm{Proposed Design} & $\approx L_{\rm max}^3 + 10L_{\rm max}^2 + 14L_{\rm max} +2$ \\
\hline \textrm{Full-State EKF} & $\mathcal{O}\left\{Q^3(L_{\rm max}N_A+1)^3\right\}$ \\
\hline \textrm{FFT-Block EKF \cite{Kalman_MIMO}} & {\tabincell{c}{$\mathcal{O}\{N_AN_F(QN_A)^2(L_{\rm max}+1)^2$\\$+QN_A(N_AN_F)^2(L_{\rm max}+1)^2$\\$+(QN_A)^3(L_{\rm max}+1)^3\}$}} \\
\hline \textrm{Parallel EKF \cite{Kalman_Particle}} & \tabincell{c}{$\mathcal{O}\{Q^2N_F(L_{\rm max}+1)^2+QN_F^2(L_{\rm max}+1)^2$\\$+QN_F(L_{\rm max}+1)^3\}$}\\
\hline \textrm{Particle Filter \cite{Kalman_Particle}} & \tabincell{c}{$\mathcal{O}\{N_PQ^2N_F(L_{\rm max}+1)^2+N_PQN_F^2(L_{\rm max}+1)^2$\\$+N_PQN_F(L_{\rm max}+1)^3\}$ \\ \textrm{\footnotesize $N_P$ is the number of particle samples.} } \\
\hline \textrm{EM method \cite{EM}} & \tabincell{c}{$\mathcal{O}\{N_LQN_F^2L_{\rm max}+N_LQN_FL_{\rm max}^2\}$ \\ \textrm{\footnotesize $N_L$ is the number of iterations.}} \\
\hline
\end{tabular}
\end{table}

\subsection{Simulation Results}
\begin{figure}
\centering
\includegraphics[height=1.75in, width=3in]{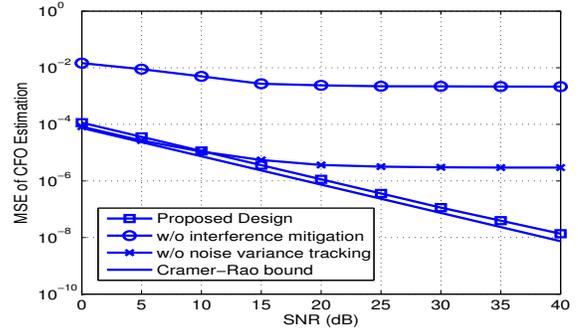}
\caption{CFO estimation's MSE results under different SNR values.} \label{MSE_CFO}
\end{figure}

\begin{figure}
\centering
\includegraphics[height=1.75in, width=3in]{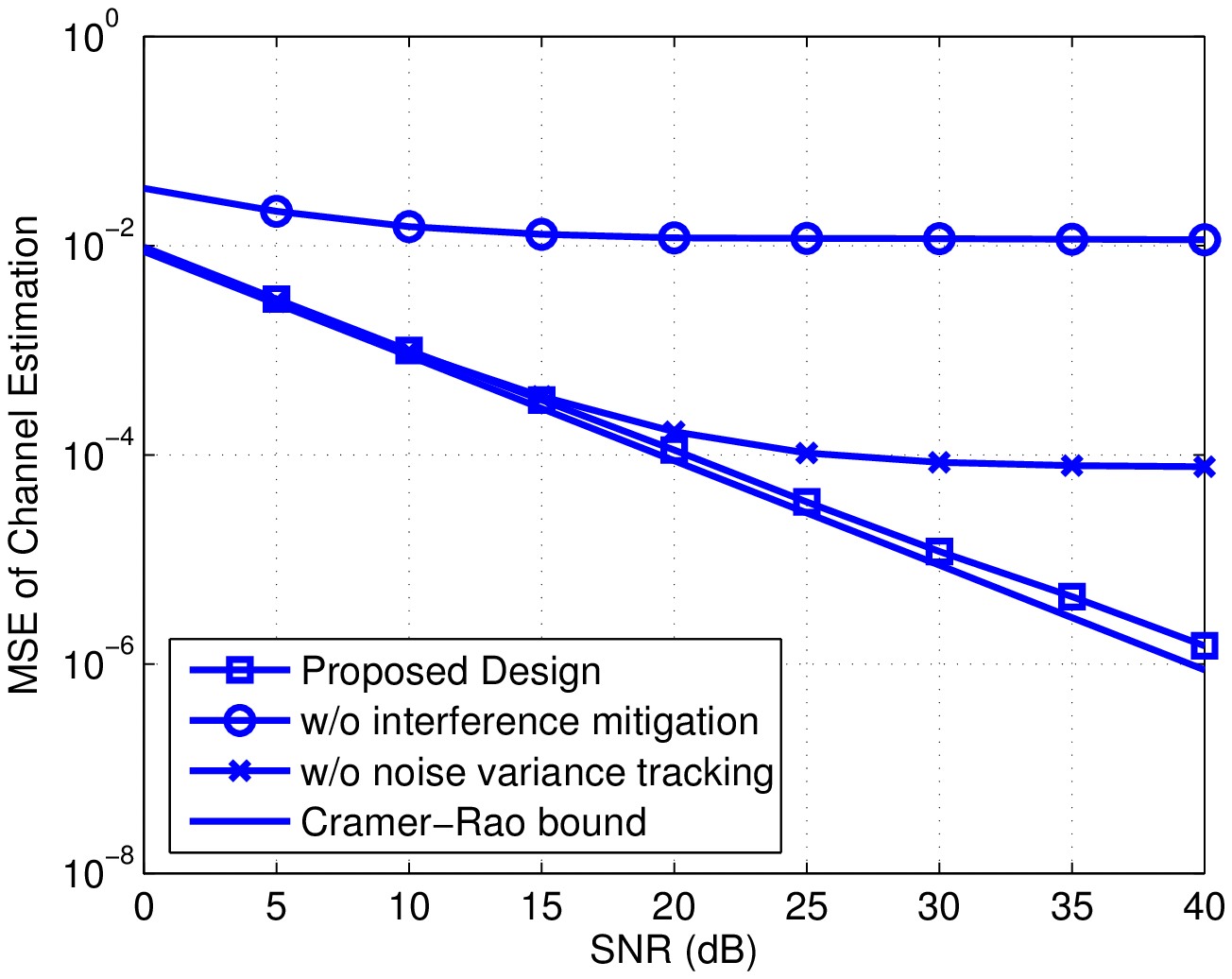}
\caption{Channel estimation's MSE results under different SNR values.} \label{MSE_Channel}
\end{figure}

We consider a cognitive network with one primary link and three secondary links (a total of 4 links). We investigate the CFO and channel estimations at one secondary RX user with $N_A=4$ RX antennas. Without loss of generality, we assume that this secondary RX user has the same received power from all TX users. We plot the MSE results of CFO and channel estimation in Fig. \ref{MSE_CFO} and Fig. \ref{MSE_Channel}, respectively. The results show that our proposed design can closely approach the Cramer-Rao bounds. In addition, we repeat our simulation by disabling interference mitigation module, or noise variance tracking module. The corresponding results (Fig. \ref{MSE_CFO} and Fig. \ref{MSE_Channel}) indicate that without interference mitigation, the estimation performance can be dramatically degraded. And without noise variance tracking module, there could be an error floor at high SNR values because of the inaccurate tracking of noise variance information. Using the values in Fig. \ref{MSE_CFO} and Fig. \ref{MSE_Channel}, it is feasible to further investigate the BER/PER performance. But such discussions heavily depend on the designed receiver structure, which is omitted here for page limitation.

\subsection{Delay, Buffer Size and Complexity Analysis}
This subsection evaluates the issues of complexity, delay and buffer size in the considered designs. We first look at the complexity issue. In particular, we count the number of complex multiplications involved in our proposed EKF design, which is listed in Table \ref{Table_complexity}. And for comparison, in that table, we also list the complexity results of Full-State EKF, FFT-Block EKF \cite{Kalman_MIMO}, Parallel EKF \cite{Kalman_Particle}, Particle filter \cite{Kalman_Particle}, and EM method \cite{EM}. Here Full-State EKF represents the EKF that builds all $(L_{\rm max}N_A+1)Q$ states in (\ref{ML}) into one state vector, yielding high state dimension. We can see that our proposed design enjoys the lowest computation complexity, which is only at the order of $L_{\rm max}^3$. But Full-State EKF's complexity is around $Q^3N_A^3$ higher than our design. Moreover, FFT-Block EKF, Parallel EKF, Particle Filter, and EM method's complexities{\footnotemark} all rely on FFT size $N_F$, which is significantly large in our case ($N_F = 2048$).

\footnotetext{Particle filters in \cite{Kalman_Particle} and EM designs in \cite{EM} have even higher complexity because of either the number of particle samples, or the number of iterations.}

Now we further look at the delay and buffer size in the proposed design. Since our EKF scheme updates the Kalman estimate at each RX sample (not at each FFT block) in an online and real-time manner, it has low estimation delay and requires low buffer size. However, Particle Filter \cite{Kalman_Particle}, Parallel EKF \cite{Kalman_Particle}, and EM approach \cite{EM} all operate at an FFT-block basis with buffer size $N_F=2048$ samples, resulting in both a large delay and a large buffer size. Even worse, particle filter and EM method both need to process the FFT-block multiple times (e.g., particle samples in particle filter, and iterations in EM method), leading to additional estimation delay.

\section{Conclusion}

This paper has presented a low-delay and low-complexity EKF design that can estimate the CFO and channel parameters in  multi-user cognitive communications. We first present a fundamental EKF design that works at one TX user and one RX antenna. Then this basic EKF design is reused in a time-shared way to estimate the parameters for multiple TX users at multiple RX antennas. Besides, an interference mitigation strategy is proposed to estimate and cancel the multi-user interference at each RX sample. Moreover, adaptive noise variance tracking module is further employed to further enhance the estimation performance. Compared with existing related designs (FFT-Block EKF \cite{Kalman_MIMO}, Parallel EKF \cite{Kalman_Particle}, Particle filter \cite{Kalman_Particle}, and EM method \cite{EM}), our proposed design enjoys low computation complexity (because of pipelining and time-sharing design), low delay and low buffer size (due to its online and run-time estimation). Besides, its estimation performance can closely approach the Cramer-Rao bound.




\bibliographystyle{IEEEtran}
\bibliography{IEEEabrv,Kalman_Filter_Globecom_final}


%




\end{document}